\documentclass[%
 aip,
 jmp,%
 amsmath,amssymb,
 reprint,%
]{revtex4-1}
\usepackage{color}
\usepackage{graphicx}
\usepackage{dcolumn}
\usepackage{bm}

\begin{document}

\preprint{AIP/123-QED}

\title[Cryogenic Millimeter-wave VPMs]{Variable-delay Polarization Modulators for Cryogenic Millimeter-wave Applications}
\author{D.T. Chuss}
\email{David.T.Chuss@nasa.gov}
\affiliation{NASA Goddard Space Flight Center, Greenbelt, MD 20771}
\author{J.R. Eimer}
\affiliation{Department of Physics and Astronomy, The Johns Hopkins University, Baltimore, MD}
\author{D.J. Fixsen}
\affiliation{NASA Goddard Space Flight Center, Greenbelt, MD 20771}
\author{J. Hinderks}
\affiliation{NASA Goddard Space Flight Center, Greenbelt, MD 20771}
\author{A.J. Kogut}
\affiliation{NASA Goddard Space Flight Center, Greenbelt, MD 20771}
\author{J. Lazear} 
\affiliation{Department of Physics and Astronomy, The Johns Hopkins University, Baltimore, MD}
\author{P. Mirel}
\affiliation{NASA Goddard Space Flight Center, Greenbelt, MD 20771}
\author{E. Switzer}
\affiliation{NASA Goddard Space Flight Center, Greenbelt, MD 20771}
\author{G.M. Voellmer}
\affiliation{NASA Goddard Space Flight Center, Greenbelt, MD 20771}
\author{E.J. Wollack}
\affiliation{NASA Goddard Space Flight Center, Greenbelt, MD 20771}

\date{\today}

\begin{abstract}
We describe the design, construction, and initial validation of the variable-delay polarization modulator (VPM) designed for the PIPER cosmic microwave background polarimeter.  The VPM modulates between linear and circular polarization by introducing a variable phase delay between orthogonal linear polarizations. Each VPM has a diameter of 39 cm and is engineered to operate in a cryogenic environment (1.5 K).  We describe the mechanical design and performance of the kinematic double-blade flexure and drive mechanism along with the construction of the high precision wire grid polarizers.  
\end{abstract}

\pacs{42,95,98}
\maketitle

\begin{quotation}
\end{quotation}

\section{\label{sec:intro}Introduction}
A Variable-delay Polarization Modulator (VPM) changes the state of polarization of electromagnetic radiation via the
introduction of a variable phase delay between two orthogonal polarization components.\cite{Chuss06}
This leads to a transfer function in which an output Stokes parameter $U^\prime$, which is defined at a $45^\circ$ angle
with respect to the polarization separation basis (defined by the VPM grid), is modulated according to 
\begin{align}
U^\prime=U\cos{\phi}+V\sin{\phi}.
\end{align}
Here, $U$ is the linear Stokes parameter that is the difference between the linear polarization components oriented at $\pm45^\circ$ with respect to the wires, $V$ is the Stokes parameter corresponding to circular polarization, and $\phi$ is the introduced phase delay between the two orthogonal polarizations.  

\begin{figure}
\includegraphics[width=2.5in]{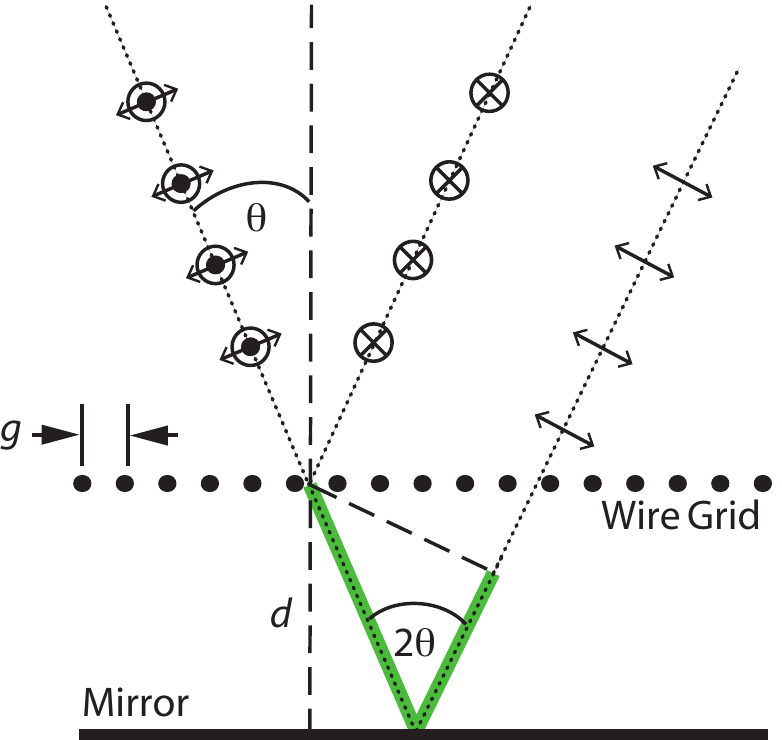}
\caption{\label{fig:dia}Overview of the VPM concept adapted from Chuss et al.\cite{Chuss12a} The geometric path difference
between the two polarizations is highlighted by the thick solid lines.}
\end{figure}

In recent work, VPMs have been realized by the arrangement of a wire grid polarizer in front of and parallel to a moving mirror \cite{Krejny08} (See Fig.~\ref{fig:dia}). This phase delay is a function of the incidence angle, $\theta$, and the grid-mirror separation, $d$. In the limit where the wavelength is much larger than the wire, the phase delay can be approximated using the geometric path difference, 
\begin{equation}
\phi\approx\frac{4\pi\lambda}{d}\cos{\theta}. 
\end{equation}
For greater fidelity, a transmission line model has been used to connect the grid-mirror separation with the introduced phase delay as a function of the wire grid geometry.\cite{Chuss12a}  In addition, metrology techniques for realizing high precision measurements of the VPM response have been developed.\cite{Eimer11}

The Primordial Inflationary Polarization ExploreR (PIPER)\cite{Chuss10,Kogut12} and the Cosmology Large Angular
Scale Surveyor (CLASS)\cite{Eimer12} will both employ VPMs as the first element of their optics in their measurement of the polarization of the cosmic microwave background on large angular scales. The motivation for using the VPM is twofold. First, the VPM can be made large enough to be placed at the primary aperture of a relevant telescope.  This allows sky polarization to be modulated while leaving any subsequent instrumental contribution to the polarization unmodulated, thereby mitigating potential mixing between temperature and polarization. In addition, the small linear distances (fraction of a wavelength) required for the grid-mirror separation variation allow a rapid ($\sim$few Hz) modulation in polarization that moves the signal out of the $1/f$ noise of the environment. 

Devices employing small linear motions have potential advantages in both reliability and power dissipation over those that utilize large angular motions for cryogenic applications. The PIPER flexures were designed for more than 3$\times10^6$ cycles, sufficient to survive 8 flights at 3 Hz operation.  It is anticipated that similar designs could be employed having much longer lifetimes.  Superconducting bearings for wave plates \cite{Klein11} are a good solution for low-friction rotational operation; however, parasitic heat removal is a challenge for such non-contacting solutions. 

This paper describes the construction and initial validation of the PIPER VPMs.  PIPER is a balloon-borne cosmic microwave background polarimeter that will operate at 4 frequencies between 200 and 600 GHz in separate flights.  The instrument is enclosed in a large bucket dewar, and each of the elements of the two telescopes is cooled to 1.5 K by a combination of evaporating liquid helium and superfluid pumps.\cite{Singal11} 
Cooling the telescope reduces background radiation and mitigates the coupling to variable grid emission.  Because of this, the VPM
must be engineered to work at 1.5 K.  

The details of the optical design \cite{Eimer10} and detectors \cite{Benford10} for PIPER are described in other papers.  This work focuses on the design, construction and initial validation of the VPMs.  

\section{\label{sec:flex}Design of the Mechanical Assembly}
\begin{figure}
\includegraphics[width=3in]{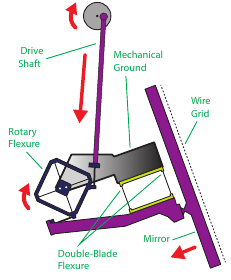}
\caption{\label{fig:schematic}The operation of the VPM drive mechanism is shown. A warm motor turns a cam shaft that is attached to a push rod via an eccentric.  This push rod turns a cold rotary flexure that in turn pushes and pulls on one side of a kinematic double-blade flexure that maintains the parallelism of the mirror over the throw. A symmetric pair of rotary flexures (not shown) couples to an additional push rod that is moved in the opposite direction of the first. These rotary flexures move counterweights to compensate the motion of the mirror.}
\end{figure}

\begin{figure}
\includegraphics[width=3.5in]{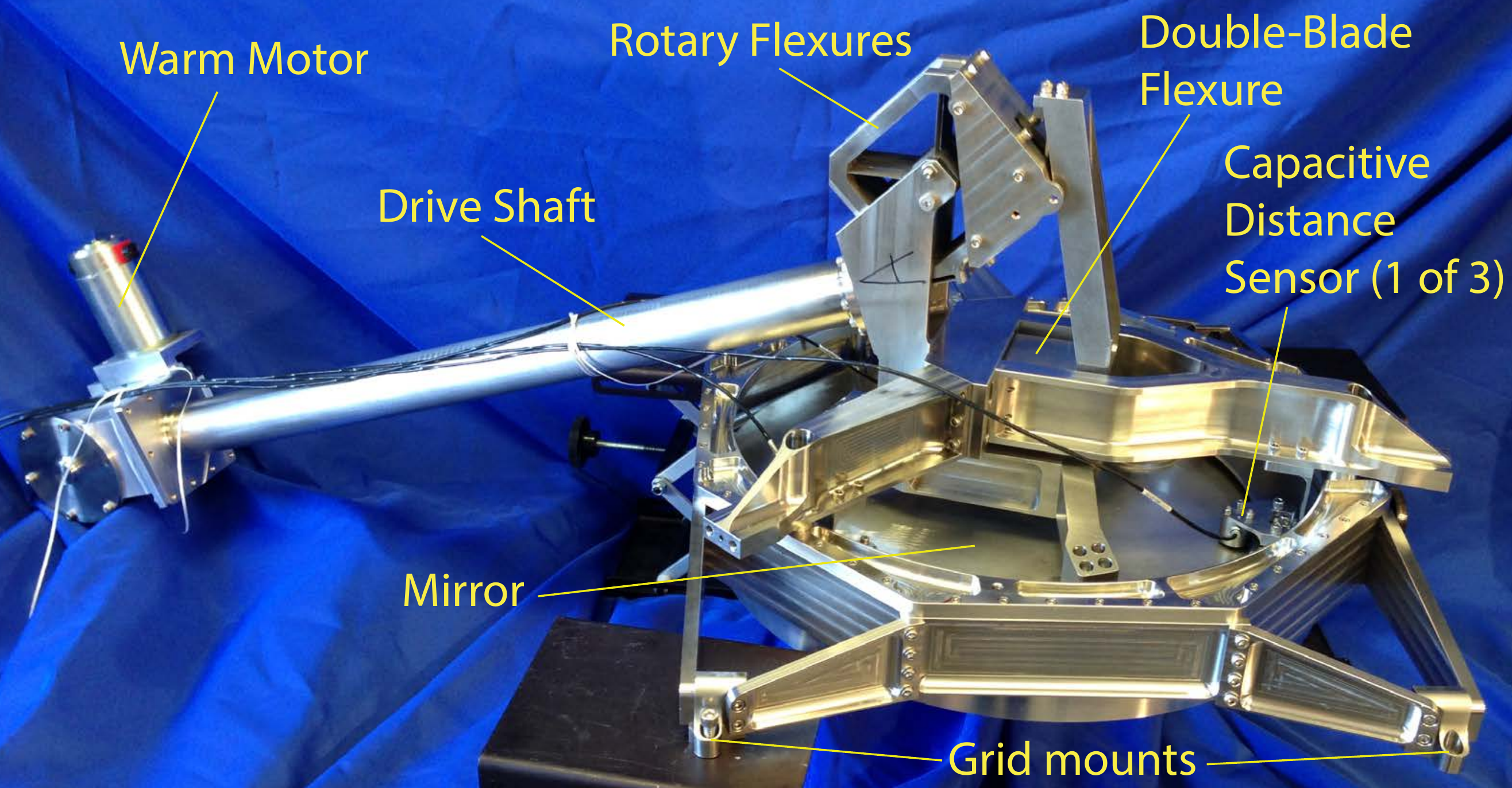}
\caption{\label{fig:vpmphoto}Photograph of the PIPER VPM assembly.}
\end{figure}

A schematic of the VPM is shown in Figure~\ref{fig:schematic}. The 1.5 K mirror is actuated by a warm (ambient temperature) motor that couples to the cold mechanism through a driveshaft. All parts of the mechanism that operate at 1.5 K are constructed from stainless steel to ensure thermomechanical stability and matching of the coefficient of thermal expansion.  This includes the mirror, which is constructed from stainless steel honeycomb.  The mirror faceplate is lapped flat, and then gold is deposited onto the mirror to increase the conductivity of the surface. A photo of the VPM drive mechanism is shown in Figure~\ref{fig:vpmphoto}.

To ensure that transport of the mirror is parallel, we have utilized a kinematic double-blade flexure similar to those used in the prototype Hertz VPM. \cite{Voellmer06}  This mechanism constrains all degrees of freedom except for the direction of motion.  The PIPER flexure is constructed out of stainless steel using wire-EDM and is larger than the Hertz flexure to permit the maximum throw required, 1 mm.  This is set by the longest wavelength modulated, $\lambda=1.5$ mm (200 GHz). The parallelism requirement of the flexure is 5 arc seconds  over the 1 mm throw to ensure that the phase is well-defined at the shortest operating wavelength, 0.5 mm (600 GHz).  In practice, the flexure is mechanically biased so that the entirety of the 1 mm throw is done on one side of the neutral (un-stressed) position.  This prevents the flexure from operating through its neutral point where its motion is potentially not as well controlled.  

The drive is powered by a motor operating at ambient temperature.  The motor drives a cam shaft. The cam shaft has two eccentrics that are 180$^\circ$ out of phase.  The primary eccentric drives a push rod that causes the central rotary flexure to rotate. This moves the mirror in a direction perpendicular to the grid; parallelism is maintained by the double blade flexure.  The secondary eccentric drives a similar mechanism that in turn drives a counterweight to compensate the mirror motion. This mitigates potential vibrations that can induce microphonic response in the system.

Because the eccentric determines the limits of the mirror movement, the operational mode of the VPM can be changed simply by replacing the cam shaft to achieve the desired throw. This allows a straightforward means to operate in different observational frequency bands.  For PIPER, the throw will be reconfigured between flights to accommodate the range of planned observing frequency bands. 

\section{\label{sec:fab}Grid Fabrication}

One of the most challenging aspects of the VPM is the construction of the polarizing grids. The grids must hold a flatness tolerance of $<$10 $\mu$m RMS over a clear aperture of 39 cm.\cite{Eimer10} In addition, to achieve high polarization diplexing efficiency, it is important to maintain high uniformity of the spacing of the wires that comprise the grid.

One historical approach for fabricating free-standing wire grid polarizers is to wind them directly onto a frame.\cite{Chambers86, Lahtinen99}  In this technique, wire spacing control is difficult because the wires are unconstrained along the frame. They can roll or slide out of position under minimal displacement forces because the wire tension, and thus the static friction holding them in place on the frame, is low during winding to avoid wire breakage.   In addition, this technique requires a single spool of wire to be used during fabrication with no tolerance to wire breakage.  An alternative to this technique that mitigates these issues is to wrap the wires on a cylinder that has been precut with regularly spaced alignment teeth, and then transfer the wires to their final frame.\cite{Novak89} The latter technique has been extended to make prototype 50 cm diameter polarizers.\cite{Voellmer08}

Here we describe significant enhancements to this technique that have led to improved wire uniformity and survivability at cryogenic temperatures. Free-standing wire grid polarizers have an advantage over comparable deposited membrane structures in that they do not use dielectric material as a substrate. The use of dielectrics causes a degradation of the capacitive mode (polarization perpendicular to wires) and decreases the maximum operating frequency for a given feature size.\cite{Whitborn85}  However, historically, the performance of free-standing wire grid polarizers has been limited due to wire spacing variability.  The grid manufacturing process presented here greatly improves the uniformity of grid wires and can therefore greatly enhance the capability of polarizers operating at wavelengths between 1 cm and 60 $\mu$m.

The process of grid fabrication is illustrated in Figure~\ref{fig:process}, and additional diagrams are included in the appendix for clarity.  The 40 $\mu$m 
copper-plated tungsten wire (36 $\mu$m diameter W wire; 0.13 $\mu$m Ni strike; 1.3-2.5 $\mu$m Cu plating)\cite{CFW} is wrapped onto a mandrel using a 5-axis CNC mill. In contrast to previous versions of this technique,\cite{Voellmer08, Novak89} the mandrel itself is smooth, with grooves cut only in two wire retainer
bars that are recessed into the mandrel. These grooves serve as guides to set the wire spacing. The precision of the milling machine used for the wrapping is sufficient that the entire mandrel need not be grooved to maintain the 1 wire per groove criterion on the wire retainer bars.  This realization reduces risk to the wire coatings that can become damaged by metal burrs during the winding and unwrapping processes.

To cut the grooves for the wires, we used a rotating end mill oriented at 45$^\circ$ with respect to the rotation axis of the mandrel.  We use a straight flute tungsten carbide tipped 1/2$^{''}$ diameter end mill
in a climb cut at 2,500 RPM on the spindle, with the mandrel rotating at 2.7 RPM.  This combination of tool speeds and feeds leaves a burr-free groove in the soft copper retaining bar. 

\begin{figure}
\includegraphics[width=3.25in]{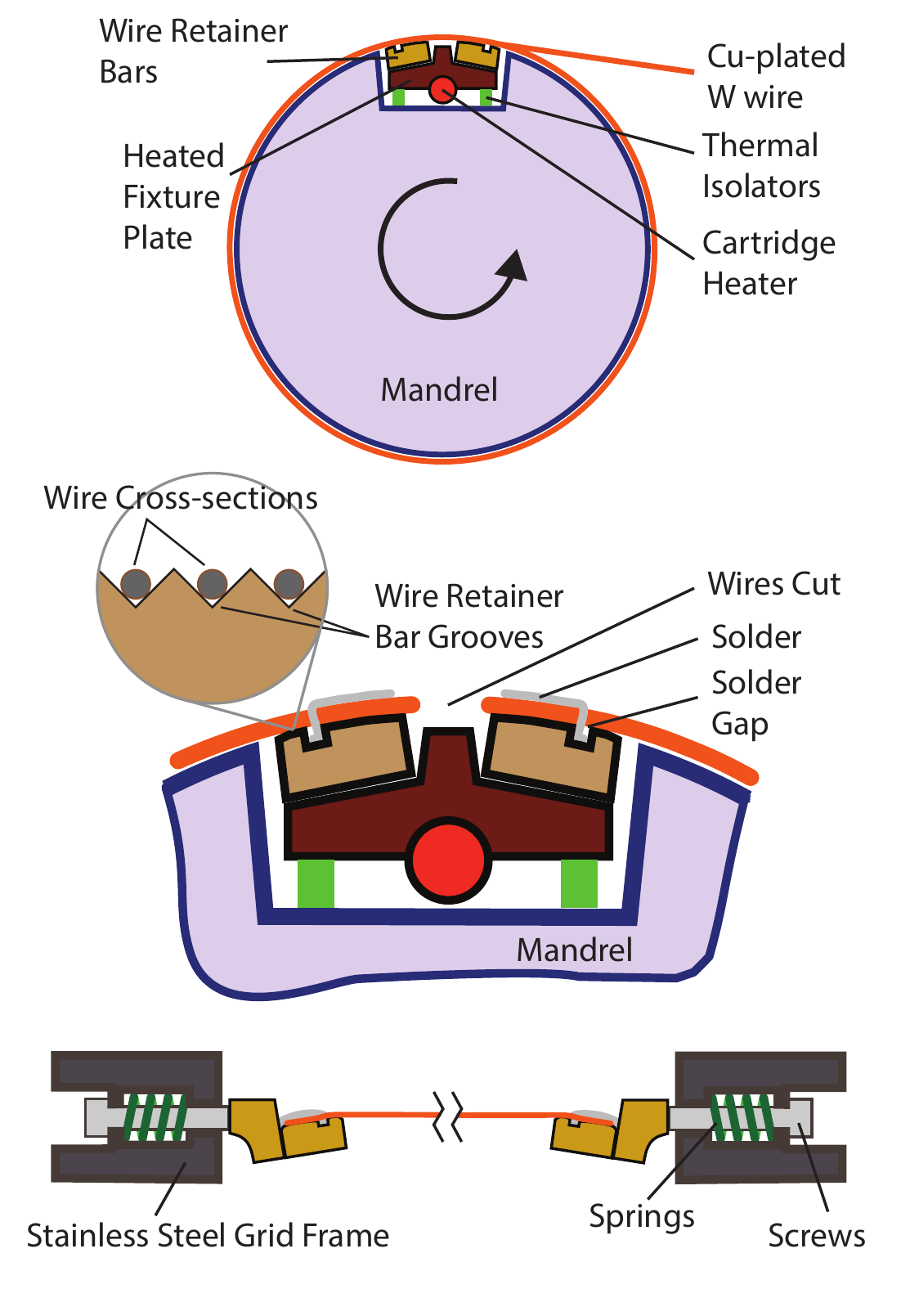}
\caption{\label{fig:process} The process for constructing the large cryogenic wire grid. (top) The wire is wrapped around a mandrel using a CNC mill. (middle) The wires are soldered to a pair of grooved wire retainer bars and cut. Heat is supplied to the wire retainer bars via an integrated cartridge heater during this process. The wire retainer bars are thermally isolated from the rest of the mandrel. (bottom) The wires are then transferred to the grid frame, attached via screws with springs. }
\end{figure}

Once the wire is wrapped, the wire is fastened to the wire retainer bars. Previous efforts have used epoxy; however, because of the large size and cryogenic operating environment of the PIPER grids, we have developed a technique for soldering the wires to the bars.  The wire retainer bars are attached to the mandrel using thermally insulating stand-offs. This enables adequate thermal isolation to allow the solder to flow on the wire retainer bars without heating the mandrel. A cartridge heater within the fixture plate heats the plate and the wire retainer bars to soldering temperature. No additional heat is added.  
The flux core solder used was manufactured by Amerway, Inc. (62\% Sn, 36\% Pb, 2\% Ag).
\cite{Amerway}  The grooves in the wire retainer bars that constrain the wire spacing are divided into two parts by a channel cut perpendicular to the wire direction.  This ``solder gap'' keeps the solder away from the clear aperture and keeps the aperture side of the grooves free of solder. This allows the wires to seat in each groove when tensioned as is explained below.

The wires are then cut at the top-dead-center location between the retainer bars and then removed from the mandrel. The retaining bars, with wires attached, are mounted to the stainless steel grid frame using screw-adjustable helical spring tensioners (See bottom of Fig.~\ref{fig:process}.) The springs maintain the tension on the wires upon cooling despite the mismatch in thermal contraction due to the dissimilar materials of the grid and grid frame. In addition, upon tensioning, the wires are pulled deeper into the unsoldered part of the retainer bar grooves.  This ensures maximum control over wire spacing. 
Additional diagrams of the mandrel and the frame are shown in Figures~\ref{fig:wdiag} and \ref{fig:winding} in the Appendix.

To define the plane of the wires, we implement a ``grid flattener,'' a polished ring pressed lightly against the wires.  This technique of separating the tensioning component from the plane definition has been implemented in previous work.\cite{Krejny08,Voellmer08}

\begin{figure}
\includegraphics[width=3.25in]{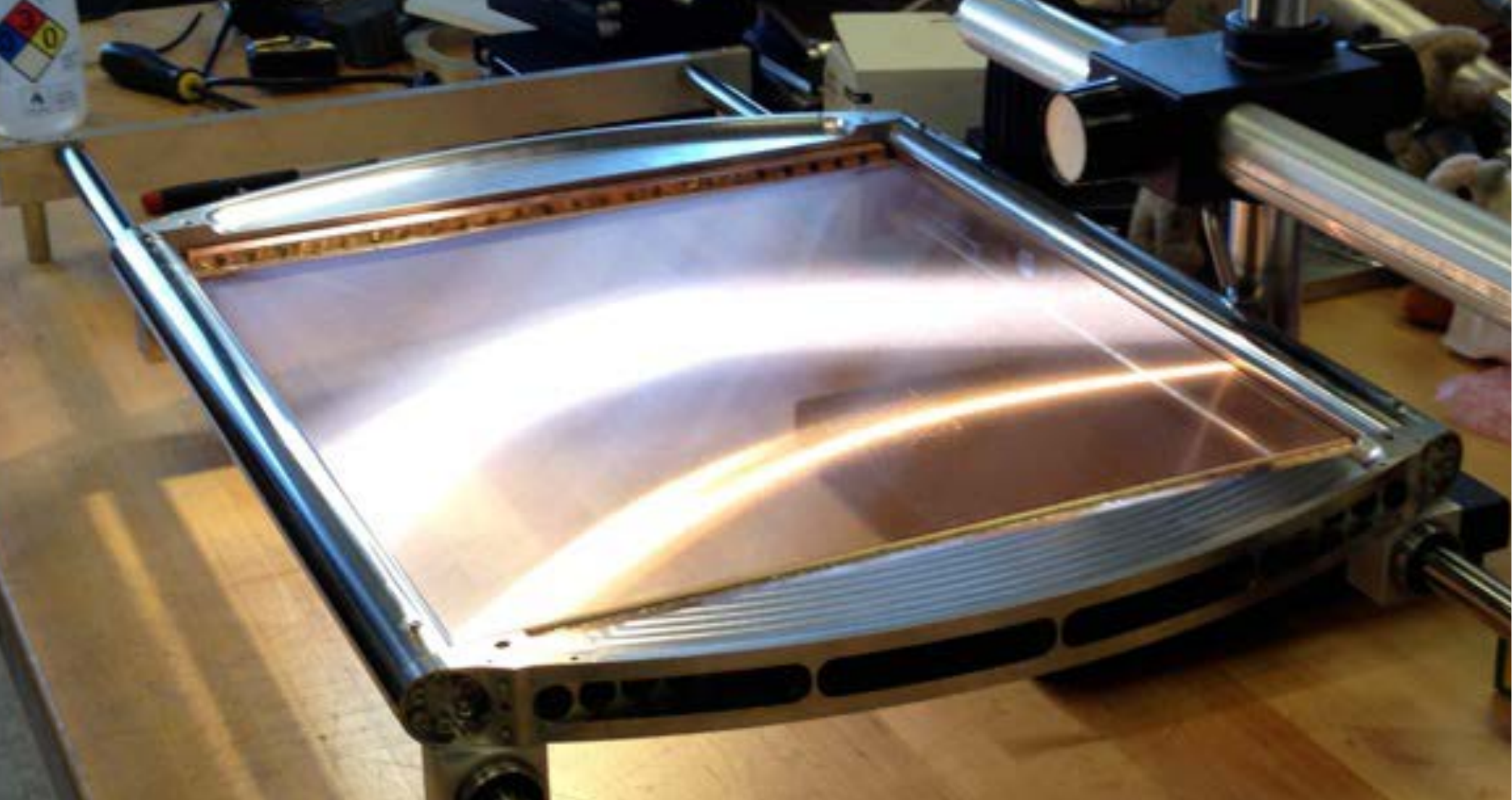}
\caption{\label{fig:grid} The final cryogenically-compatible polarizing grid is shown.  The dimensions of the grid are 41.7$\times$40 cm$^2$.}
\end{figure}

The tension is set to be 0.7 N (130 MPa) on each wire, or approximately 25\% of the breaking strength of the tungsten wire. This is implemented by measuring the spring constants and setting the compression distance.  This leads to a resonant frequency of 190 Hz, before application of a grid flattener. Application of the grid flattener increases the resonant frequencies of the individual wires.
An image of the final grid is shown in Figure~\ref{fig:grid} before application of the flattener. 

To summarize, we highlight key innovations in this fabrication process that separate it from previous work.\cite{Novak89, Voellmer08}
(1) The grooves have been cut on the retainer bar only. (2) The wire-guide grooves are cut using an end mill oriented at 45$^\circ$ to the mandrel axis to produce highly regular and burr-free cuts. 
(3) The wires are soldered to the retainer bars for cryogenic compatibility using an {\emph {in situ}} soldering technique. 
(4) The interface between the retainer bars and the grid frame is designed to pull the wires deeper into the grooves that ultimately define the spacing uniformity. 
(5) The entire wire retainer bar is attached to the grid frame, so the wires are put into their position-defining grooves once and never removed. 

\section{\label{sec:validation}Validation}
\begin{figure}
\includegraphics[width=3.25in]{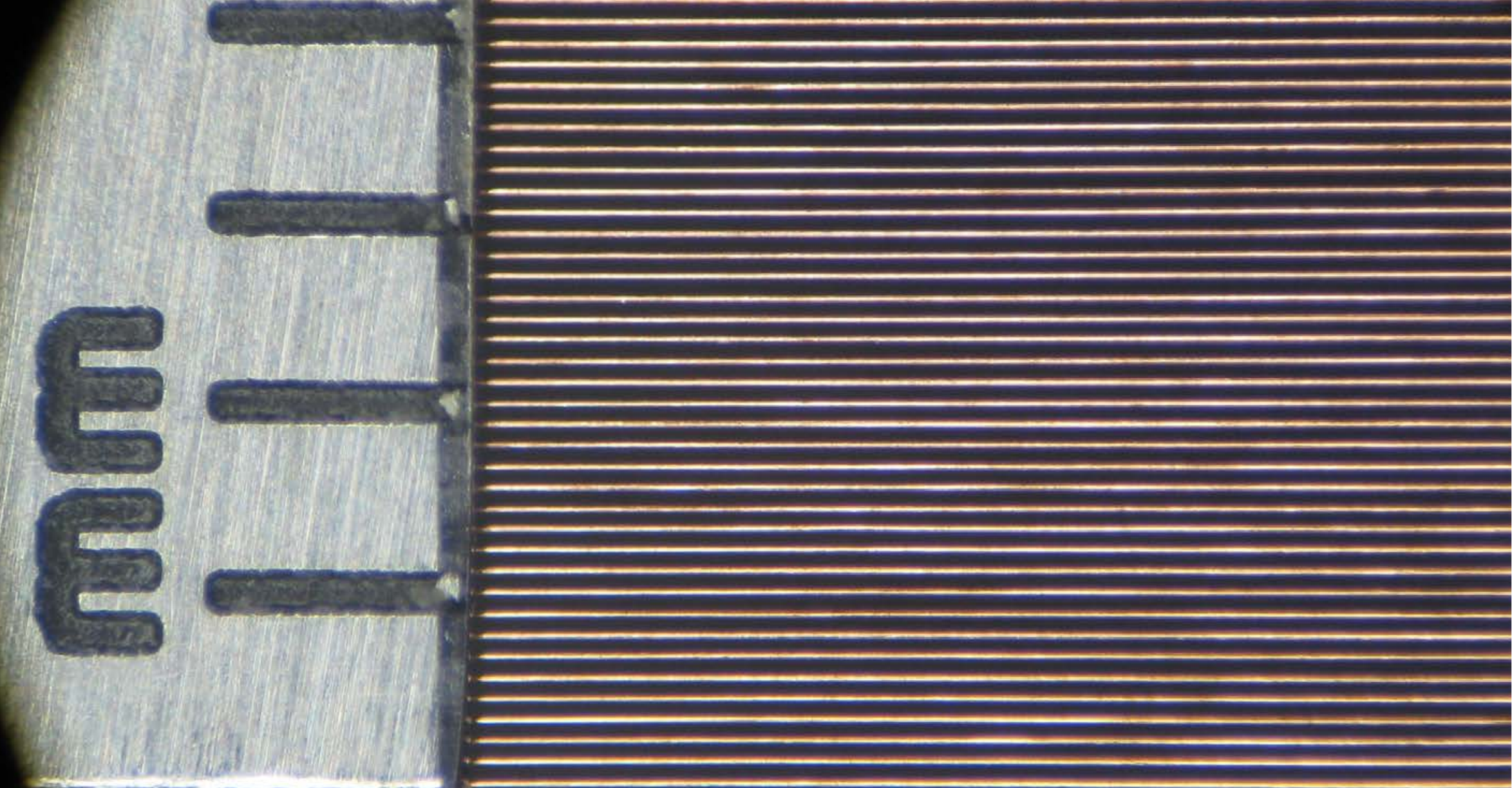}
\caption{\label{fig:wires} A micrograph of a section of the wires is shown with a millimeter scale for comparison. The spacing of the wires is $117.0\pm 5.7 \,\mu$m. }
\end{figure}

\subsection{Uniformity of Wire Spacing}
The wire spacing has been measured using a STIL S.A. CHR 150 confocal spectrometer.\cite{STIL} In this configuration, the grid was placed on the X-Y stage of a milling machine, and the vertical distance to the grid wires was measured as a function of horizontal position.  The average spacing of the wires was found to be $117.0\pm 5.7 \,\mu$m. The error is determined by the standard deviation of many measurements taken using the confocal spectrometer. The reported error is partially due to the finite beam size of the confocal spectrometer beam that is not well characterized.  Thus, this measurement represents an upper limit on the 1-$\sigma$ deviation.  A micrograph of the wires in Figure~\ref{fig:wires} shows the wire uniformity.

\subsection{Flatness and Parallelism}
\begin{figure}
\includegraphics[width=3.25in]{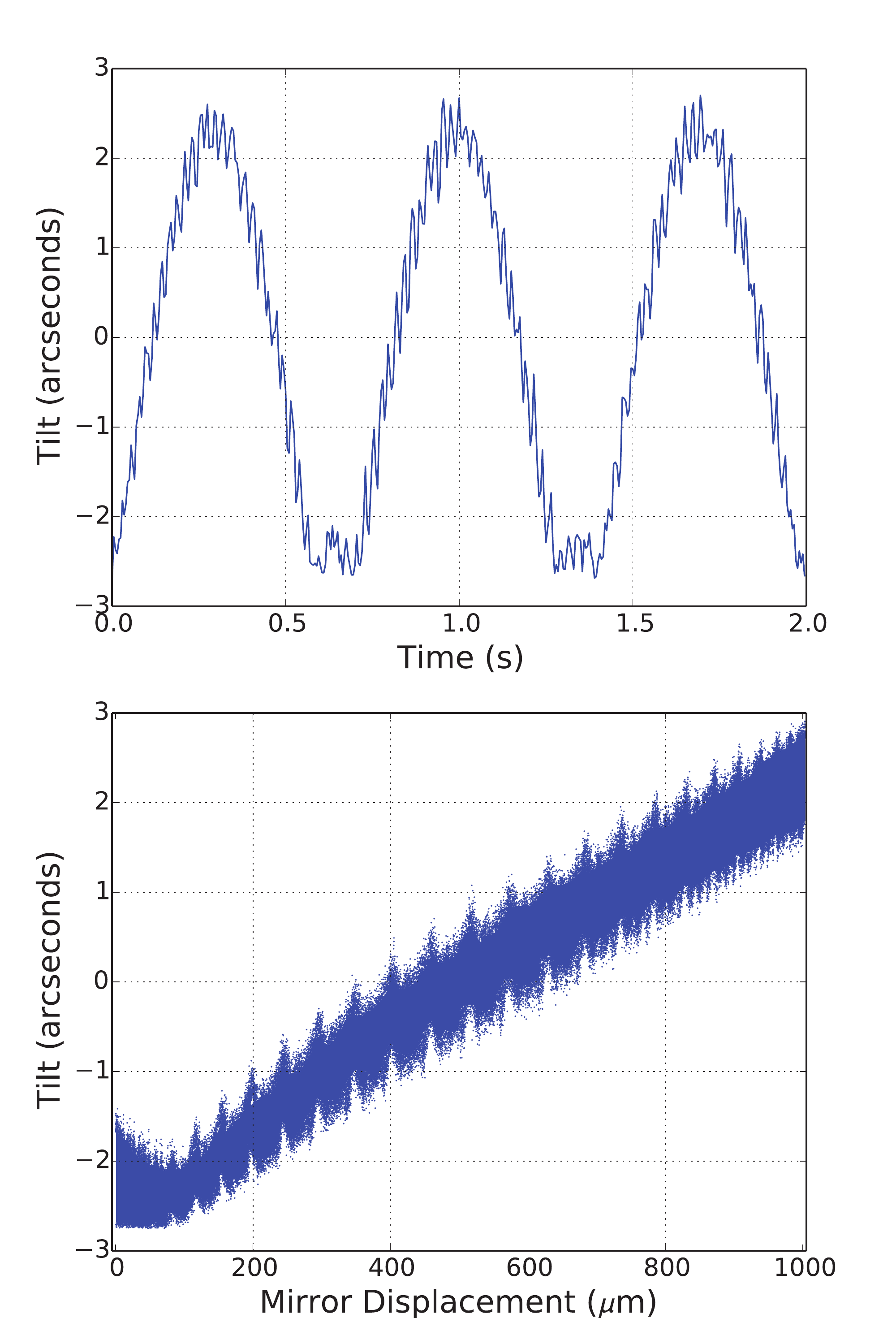}
\caption{\label{fig:par} (Top) The mirror tilt is shown as a function of time. Here the cam is driven at a frequency of 1.4 Hz using a test motor. Several periods of the motion are shown.  (Bottom) The tilt is shown as a function of the mirror displacement a for 10 minutes of data. At each point, the mirror displacement is determined by averaging the readings from the three capacitance sensors. }
\end{figure}

The grid-mirror separation is measured at three places near the edge of the mirror, evenly spaced in angle 120$^\circ$ from each other.  Capacitance sensors\cite{KLA} are mounted to the frame and measure the distance between the sensor and the back of the mirror with better than 0.5 $\mu$m precision using a standard calibration curve.  Figure~\ref{fig:par} shows the tilt of the mirror as a function of time for a few modulation cycles (top) and mirror displacement for 10 minutes of data (bottom).  The tilt was calculated using the displacement data from the three capacitance sensors. The deviation from parallelism is $\sim\pm2.5$ arc seconds. This corresponds to a maximum tilt of about 7 $\mu$m across the 30 cm illuminated aperture of PIPER.  At 270 $\mu$m, this is 0.5\% of the wavelength, and so the induced phase error is small. 

 The grid flatness with the current grid flattener is $\pm 8.7 \mu$m (1-$\sigma$) as measured with the confocal spectrometer.


\subsection{Cryogenic Compatibility}
A test grid was constructed by wrapping the central 15 cm of the PIPER grid frame at the full 41.7 cm length.  This sample was cycled to 77 K ten times and did not fail. 
In addition, the flexure mechanism and drive train were cooled to 4 K in a liquid helium dewar and remained fully operable. 

The VPM mirrors are constructed of stainless steel honeycomb that is internally held together using a brazing process developed by Bodycote.\cite{Bodycote}  The face sheet used for the mirror was 1.5 mm material that will be lapped to a flatness of $<$5 $\mu$m and vapor deposited with 2.5 $\mu$m of gold.  We tested this construction for thermal survivability by thermally cycling the mirror 10 times to liquid nitrogen (77 K) temperature.  We found no change in the flatness; however, the precision surface (lapped and vapor deposited) was not applied.

\subsection{Grid Polarization Efficiency}
\begin{figure}
\includegraphics[width=3.0in]{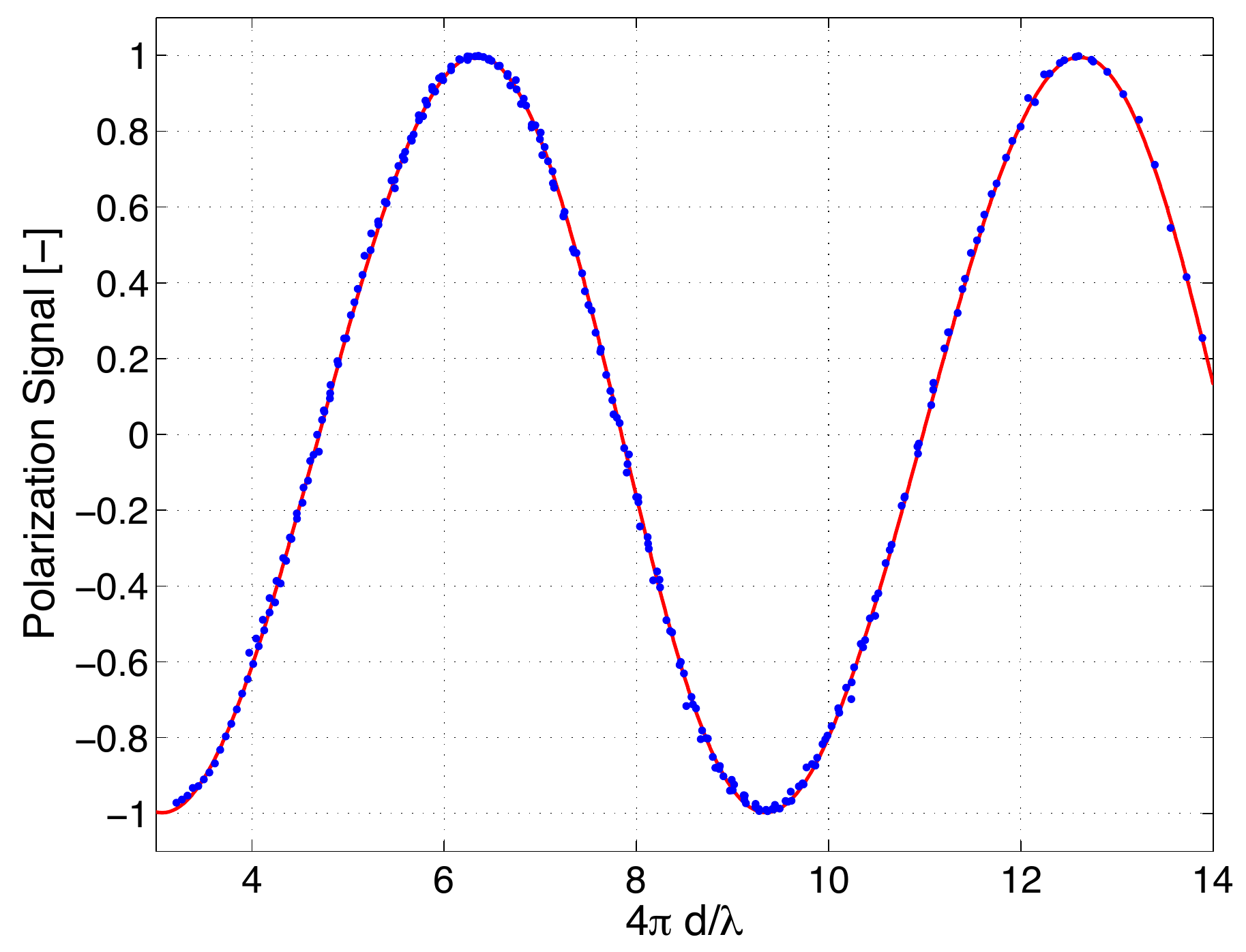}
\caption{\label{fig:efficiency} The polarization transfer function of a sample grid in a test VPM is shown.  The data points are polarization measurements as a function of phase from 76-117 GHz.  The measurement is consistent with unit efficiency to the accuracy of the measurement. The red curve is a transmission line model\cite{Chuss12a} evaluated for the measured grid parameters.}
\end{figure}

We have made initial measurements of the grid efficiency using a quasioptical testbed coupled to a Vector Network Analyzer (VNA).  We have transferred a sample of a test grid to a 15 cm diameter circular frame and mounted it to a small VPM. Measurements of the polarization transfer function were made from 76-117 GHz. The data analysis is identical to that for previously published results.\cite{Chuss2012b} Figure~\ref{fig:efficiency} shows the single frequency polarization transfer function from 76-117 GHz. For each grid-mirror separation, a polarization spectrum is obtained. Each point in Figure~\ref{fig:efficiency} is the mean of 100 data points (1.36 GHz bandwidth.) Data from 9 grid-mirror separations are included in the plot. 

For reference, we have superposed the transmission line model for a VPM from Chuss et al.\cite{Chuss12a} evaluated using the mean wavelength over the band (3.15 mm) and the PIPER wire diameter (40 $\mu$m.)  The resistive circuit element has been calculated using the conductivity of copper. The measured grid efficiency is greater than 99\%. This measurement is limited by the uncertainty in the VNA absolute calibration.

\section{Summary}
We have designed, constructed, and validated a cryogenic variable-delay polarization modulator (VPM) for the PIPER suborbital cosmic microwave background polarimeter. The achieved specifications for the VPM are shown in Table~\ref{tab:vpmsum}.

\begin{table}[htbp]
   \centering
   \begin{tabular}{@{} lcc @{}} 
      \toprule
      Property   & Value & Units \\ \hline
     Maximum mirror throw      & 1.0 & mm \\
     Mirror tilt at maximum throw          & 5     &  arc seconds \\
     Clear aperture	& 39 	& cm\\
     Wire diameter       & 40  & $\mu$m \\
     Wire separation       & 117.0  & $\mu$m \\
     Wire separation error & 5.7 & $\mu$m\\
     Grid flatness & 8.7 & $\mu$m\\
      Min. wire resonance & 190   &  Hz \\
      Polarization Efficiency & $>99$ & \%\\
      \hline
   \end{tabular}
   \caption{The parameters of the VPM for PIPER.}
   \label{tab:vpmsum}
\end{table}
\section*{Acknowledgements}
We thank Alyssa Barlis, Adam Blake, Matheus Teixeira, and Vien Ha for work in the initial testing of the fabrication process and Paul Cursey for machining support. We thank Mackenzie Turvey for assistance with the electromagnetic testing. In addition, we would like to thank Mike Jackson for modeling and finite element analysis for the flexure and drive system. This work was funded by a NASA APRA suborbital grant.

%
\section*{Appendix: Diagrams of Mandrel and Grid Frames}
\begin{figure}[h!]
\includegraphics[width=3.25in]{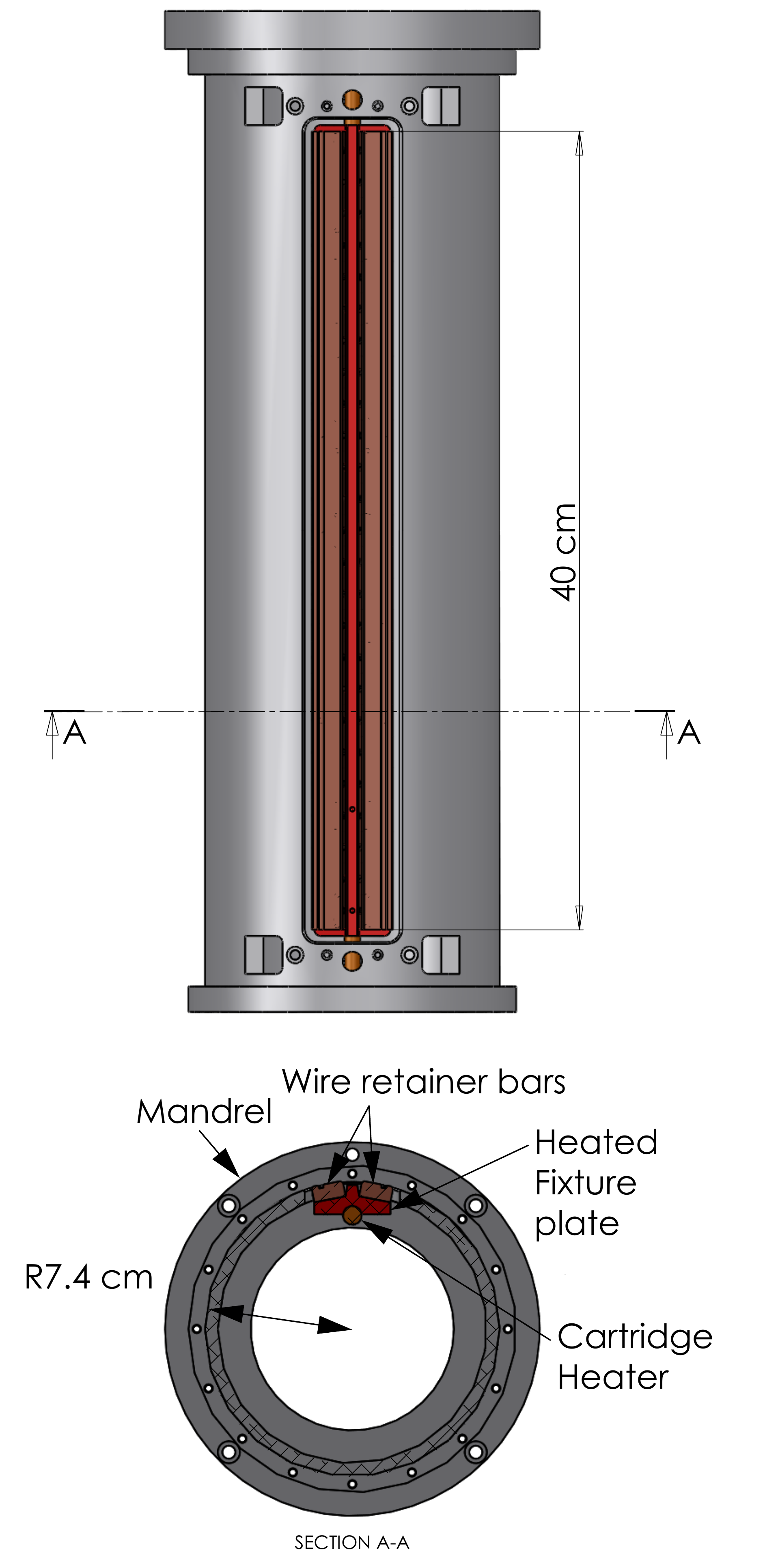}
\caption{\label{fig:wdiag} The mandrel for winding the grids is shown. The wire is wrapped around a mandrel using a CNC mill. The wires are soldered to a pair of grooved wire retainer bars and cut. The wires are then transferred to the grid frame, attached via screws with springs. }
\end{figure}

\begin{figure}[h!]
\includegraphics[width=3.25in]{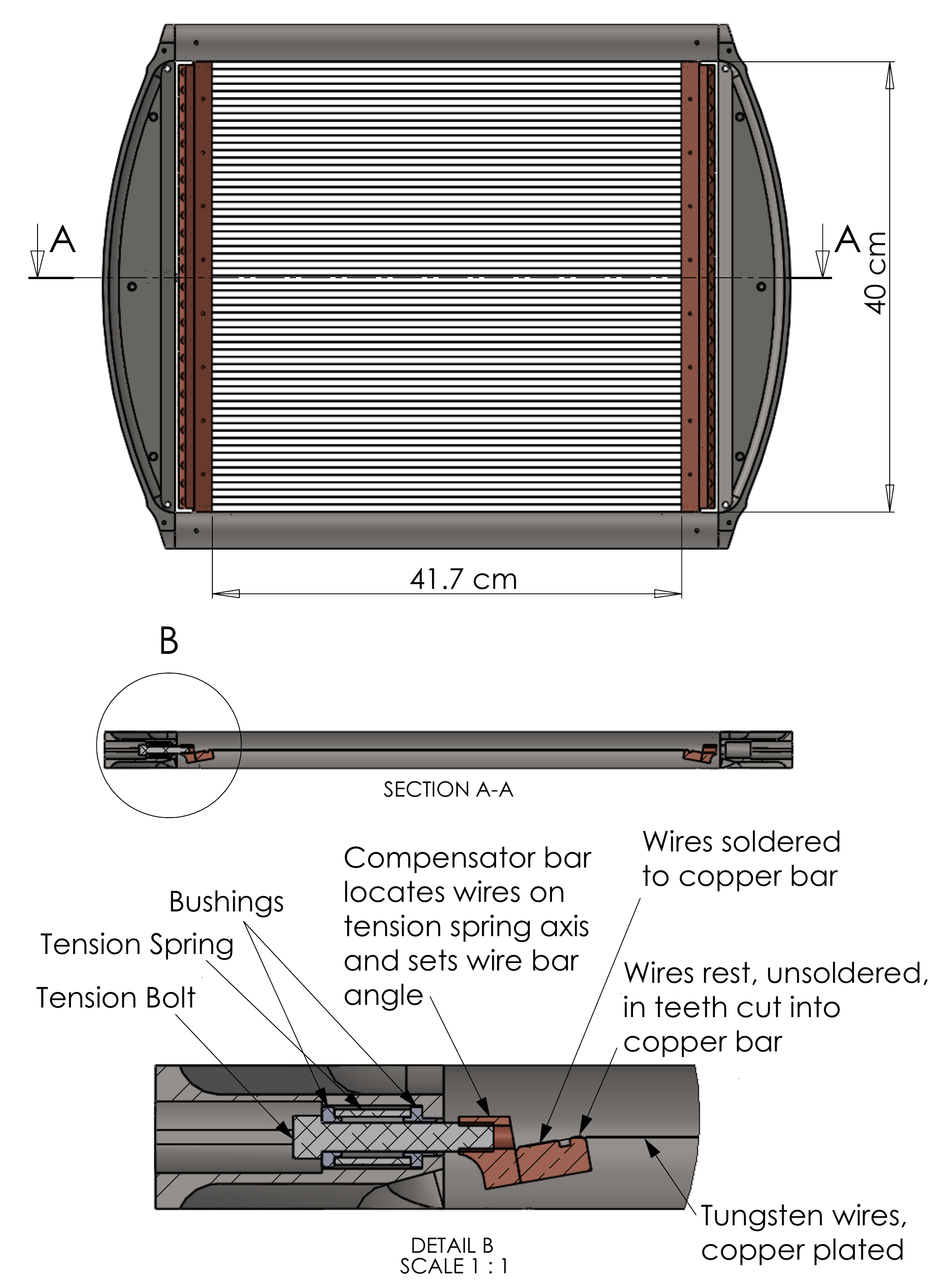}
\caption{\label{fig:winding} Two views of the grid are shown (top and middle) along with a detail drawing of the attachment of the wire retainer bars (bottom). }
\end{figure}

\end{document}